# Design and Focused Ion Beam Fabrication of Single Crystal Diamond Nanobeam Cavities

Thomas M. Babinec[1], Jennifer T. Choy[1], Kirsten J. M. Smith[1,2], Mughees Khan[1,3], Marko Lončar[1]

[1] School of Engineering and Applied Sciences, Harvard University, Cambridge, MA 02138, U.S.A.

[2] Department of Physics and Astronomy, Vrije Universiteit, The Netherlands

[3] Wyss Institute, Harvard University, Boston, MA 02115, U.S.A.

**Abstract**

We present the design and fabrication of nanobeam photonic crystal cavities in single crystal diamond for applications in cavity quantum electrodynamics. First, we describe three-dimensional finite-difference time-domain simulations of a high quality factor ($Q \sim 10^6$) and small mode volume ($V \sim 0.5\ (\lambda/n)^3$) device whose cavity resonance corresponds to the zero-phonon transition (637nm) of the Nitrogen-Vacancy (NV) color center in diamond. This high Q/V structure, which would allow for strong light-matter interaction, is achieved by gradually tapering the size of the photonic crystal holes between the defect center and mirror regions of the nanobeam. Next, we demonstrate two different focused ion beam (FIB) fabrication strategies to generate thin diamond membranes and nanobeam photonic crystal resonators from a bulk crystal. These approaches include a diamond crystal "side-milling" procedure as well as an application of the "lift-off" technique used in TEM sample preparation. Finally, we discuss certain aspects of the FIB fabrication routine that are a challenge to the realization of the high-Q/V designs.

## I. Introduction

Single crystal diamond is an exceptional material for future nanoscale mechanical, electronic, and photonic systems[1]. A specific example of a diamond device that would have a broad impact is an optical resonator possessing a long photon lifetime (high quality factor Q) as well as strong light confinement (small mode volume V). For example, biological and chemical sensing schemes based on optical resonators[2-3] could take advantage of the biocompatibility of diamond. The material hardness and high Young's modulus of diamond could be utilized in high-frequency opto-mechanical systems[4-6]. Finally, engineered light-matter interactions between individual defect centers in diamond[7-13] and a resonator would enable new explorations in cavity quantum electrodynamics[14-15] and quantum information processing at room temperature.

Before such applications are realized, however, substantial progress needs to be made in the design and fabrication of single crystal diamond devices. The diamond material system presents two main challenges to the realization of high Q/V nanocavities. First, light confinement is generated via processes that rely on the contrast in refractive index between the device material and surrounding medium. The modest refractive index of diamond (n ≈ 2.4) relative to air narrows the photonic bandgap of traditional two-dimensional photonic crystals and weakens the localization of the cavity mode. As a result, large and complex heterostructure designs have been required to boost cavity performance beyond the level of $Q \sim 10^4$ and $V \sim (\lambda/n)^3$ [16-19], where $\lambda$ is the resonant wavelength and n is the refractive index of diamond. Second, commercially available single crystal diamond samples are presently available only in bulk form. This is a significant practical obstacle to the generation of freestanding, single-mode nanocavities using standard planar fabrication tools such as reactive ion etching and electron beam lithography.

In this letter we address each of these issues and describe a novel theoretical and fabrication tool for diamond nanophotonics. In section II we present the design of a high quality factor (Q ~ $10^6$) and small mode volume (V ~ 0.5 $(\lambda/n)^3$) single crystal diamond nanobeam photonic crystal resonant to the zero-phonon line (637nm) of the Nitrogen-Vacancy center in diamond. This design, which is based on recent theoretical[20-24] and experimental[25-28] progress implementing nanobeam cavities in other material systems, is sufficient to realize the strong-coupling regime for a Nitrogen-Vacancy center (based on recent analysis performed elsewhere[23]). In section III we investigate the suitability of the focused ion beam (FIB) milling tool to realize this device. In particular, we implement two focused ion beam (FIB) machining strategies, which are based on traditional sample preparation techniques in transmission electron microscopy[29-31], to fabricate free-standing diamond membranes. These are subsequently patterned with a waveguide and one-dimensional array of holes to generate nanobeam photonic crystals. However, milling conditions with the FIB yielded rounded sidewalls, one-of-a-kind devices, and other fabrication imperfections that present a challenge for realizing high Q/V structures. Finally, we conclude in section IV with an outlook for alternatives to this approach that could be pursued in the future.

## II. Design of High Q/V Nanobeam Cavities in Diamond

The following is a general description of the procedure used to design a diamond nanobeam photonic crystal. The nanobeam photonic crystal device consists of a suspended diamond waveguide punctured with a regular one-dimensional lattice of air holes that defines a periodic dielectric profile. An optical cavity and localized mode are formed in the nanobeam by introducing a small shift in the relative position of two holes in the linear array (Fig. 1a). Three-dimensional light confinement is then generated by Bragg reflection along the length of the waveguide and by total internal reflection in the two transverse directions. Recent work has shown that additional engineering of the dielectric profile can suppress radiation losses at the interface between the cavity and mirror regions [23,32-34] if the effective index of the photonic crystal waveguide mode $n_{wg}$ matches that of the Bloch mode $n_{Bl} = \lambda / 2a_0$ in the mirrors. In practice, this can be achieved by adiabatically increasing the periodicity *a* of the holes away from the cavity center while keeping the hole radius *r* and width *w* of the beam fixed with respect to *a*. The result is a dramatic increase in the mirror reflectivity and corresponding cavity quality factor Q with only small change to the overall mode volume V. The resonance wavelength $\lambda$ of the nanocavity can then be tuned via the cavity length parameter *s*, which is defined as the gap between the two center holes in the cavity. For each choice of mirror periodicity *a*, there is an optimal *s* at which the confinement of the cavity mode is the strongest and Q is maximized.

We now present a concrete example of one of our air-bridge diamond nanobeam cavity designs that was modeled using a three-dimensional finite-difference time-domain (FDTD) solver (RSOFT). The thickness and width of the diamond nanobeam were set to 150 nm and 264 nm, respectively, which support a single transverse electric (TE) mode. These parameters were somewhat arbitrary and can be modified depending on sample conditions in the future. The ratio of the hole radius to periodicity was fixed at $r = 0.28a_0$ in the mirror section. For the range of $a_0 \sim$ 220-230 nm, this resulted in a wide photonic bandgap with ~20% gap-to-midgap ratio at red wavelengths. Finally, we considered a nanobeam cavity that consisted of 15 mirror pairs and a 5-hole linear taper from $a_0 = 225$nm in the mirror secion to $a_5 = 179$nm at the cavity center (Fig.1b, inset). As the cavity length $s$ was scanned over the range 70 to 95nm (Fig. 1b), a maximum quality factor of Q = 3.6 x $10^6$ is achieved at a mode volume of V = 0.45$(\lambda/n)^3$. The resonant wavelength for optimal cavity lengths s ~ 80-85nm, where the quality factor is maximized at Q ~ $10^5$-$10^6$, is between 635 and 638 nm. This provides optimal spectral overlap between the resonator and the zero phonon line (637 nm) of the NV center in diamond.

This device platform has several benefits for future photonic quantum information processing systems based on diamond. First, the nanobeam cavity provides a high Q/V ratio in a small device footprint, which could be useful since thin film material is scarce. Second, the nanobeam cavity may be integrated with waveguides[26,35] for distributing single photons on- and off-chip. Finally, we note that the nanobeam design parameters presented here may be scaled in order to change the resonance wavelength[36] to overlap with the emission spectrum of other color centers in diamond besides the NV center. One obvious challenge, however, is that the cavity Q and resonant wavelength are extremely sensitive to nanometer-scale shifts in the positions and dimensions of the holes. This highlights the strict fabrication tolerances that are required for the development of a high Q/V structure.

### III. FIB Machining of Nanobeam Photonic Crystal Devices

The major challenge for the implementation of this device is the generation of freestanding diamond membranes that may be patterned with waveguides and a photonic crystal lattice. Research has thus far focused on using ion-slicing membranes from a bulk diamond crystal[37-40], which is analogous to the "smart-cut" process used to generate planar devices in silicon photonics[41]. Here we demonstrate how three-dimensional focused ion beam (FIB) milling techniques can be used to make free-standing photonic crystal nanobeam cavities directly in a bulk diamond crystal. These fabrication procedures are advantageous for applications to diamond, where individual color centers could be integrated into a nanophotonic structure via ion implantation and thermal annealing with nanometer-scale precision. Ion milling with the FIB also overcomes fabrication challenges associated with the mechanical hardness of diamond.

The diamond sample was machined using a FEI Strata 235D FIB. An energetic 30keV beam of Gallium ions with programmable 1pA-30nA/cm$^2$ current was rastered across the sample at a pixel dwell time of 1μs and at 50% pixel overlap, which caused the irradiated diamond to be sputtered (Fig. 2). The diamond sample was approximately 3 x 3 x 2 mm$^3$ in size, and was mounted on a five-axis stage that allowed for in-situ rotation of the sample. Direct writing of structures may be performed with a resolution of ~5nm at 1pA/cm$^2$ beam current, though this was limited in practice by a Gaussian beam profile whose tail broadens at higher beam currents. Consequently, there was a trade-off between the amount of time required to perform each milling step and the resolution of the corresponding features patterned in the diamond. Features with progressively smaller sizes were milled at decreasing beam current in a series of fine polishing stepse. This allowed for a device to be fabricated in a reasonable amount of time (hours).

One approach that we have implemented to generate diamond membranes and a nanobeam photonic crystal was based on a crystal 'side-mill' procedure. This technique relies upon the FIB's in-situ sample rotation in order to perform three-dimensional sculpting (Fig. 3a). First, the diamond sample was placed in a 45° mount, and rotated so that the ion beam was square with respect to the side edge of the diamond sample. Second, a large, square pit that was ~15μm long, 3μm high, and ~3μm deep was milled into the diamond crystal using ~700-1000pA/cm$^2$ beam current. This released a ~1μm thick diamond membrane whose top side corresponded to the surface of the bulk diamond crystal and whose bottom side corresponded to the top of the milled pit (Fig 3b). A polishing procedure was then used to reduce the thickness of the membrane from ~1μm down to ~150-200nm by milling the underside of the membrane with progressively lower beam currents (~150-300pA/cm$^2$). Next, the sample was rotated so that the top surface of the diamond membrane was square with respect to the ion beam. Finally, the membrane was patterned at ~10-40pA/cm$^2$ with long rectangles in order to realize an array of waveguides. These were then milled with holes to generate the photonic crystal mirror and cavity (Figs. 3d-e). Significant re-sputtering of material on the bottom of the nanobeam cavities was observed during this last step (Fig. 3e, light-gray layer), though it could be removed using strong acid cleaning in the future.

An alternative approach to generate nanobeam cavities was based on the traditional 'lift-out' procedure used for TEM sample preparation[29-31]. The first step was to define a 1-2μm thick, vertical ridge in the top side of a diamond crystal using a ~1000pA/cm$^2$ milling step (Fig. 4a). A microprobe was then attached to the top side of the diamond ridge using ion beam assisted chemical vapor deposition of tungsten at the joint. The left and right side of the ridge were then milled through to the bottom of the pit. Rotating the sample to ~30º allowed us to mill through the underside of the membrane and to completely release it from the bulk diamond sample. This thin diamond membrane was then transferred and welded to a TEM grid so that it was suspended in air (Figs. 4b-c). At this point the slab was then released from the microprobe by further milling and polished in order to remove any residual metal layer that may have been deposited on the diamond. The membrane thickness was thinned to ~150-200nm and patterned with waveguides and nanobeam cavities using a similar approach (Fig. 4d-e).

## IV. Conclusions and Future Directions

We have observed that the FIB is a flexible and versatile tool to perform three-dimensional sculpting of nanobeam photonic crystal structures from a bulk diamond crystal. There are still several challenges to the successful implementation of this technique in high Q/V designs such as the one described here. Substantial rounding of critical nanobeam cavity features, such as in the definition of the waveguide and in the patterning of the linear array of holes, results from the wide FIB beam that extends beyond the intended direct-write area. This presents a major obstacle to the fabrication of high-Q photonic crystal cavities that are exponentially sensitive to cavity spacing at the ~1nm level. Moreover, realization of these complex photonic structures is time consuming and few structures can be made within a reasonable amount of time due to the serial nature of the FIB-based fabrication. This limits probability of: (i) having a NV center within a given cavity, unless sophisticated implantation approaches are used[42]; (ii) achieving the high Q/V design for any resonance wavelength; and finally (iii) achieving cavity resonance at the zero-phonon line wavelength of the NV center without implementing an additional tuning mechanism.We therefore expect that, with this purely FIB-based fabrication approach to generate the nanobeam cavity, it will be challenging to achieve strong light-matter interactions with a NV center.

One approach that could be used in the future is to utilize the FIB in less-complex photonic structures. For example, we have fabricated both diamond nanowire waveguides as well as solid-immersion lenses using two-dimensional FIB patterning techniques (images not shown)[43].

This approach is far simpler to implement than complex three-dimensional sculpting, and these devices have already been shown to be useful for increasing the number of single photons collected from an individual NV center[44-45]. Another strategy would be to substitute other diamond nanofabrication techniques[39,46-47] such as lithography and reactive ion etching wherever possible in the "lift-off" and "side-milling" schemes presented here. This could result in greater device yield and improved device quality. Finally, devices based on thin single crystal diamond films promise many exciting applications and would likely offer the greatest opportunity for fabricating high quality diamond photonic systems.

**Acknowledgements**

Devices were fabricated in the Center for Nanoscale Systems (CNS) at Harvard. We thank Richard Schalek and Fettah Kosar for many helpful discussions and FIB assistance, and Daniel Twitchen from Element Six for helpful discussions and for diamond samples. T. M. Babinec was supported by the NDSEG fellowship, and J. T. Choy by the NSF graduate student fellowship. This work was supported in part by Harvard's Nanoscale Science and Engineering Center (NSEC), NSF NIRT grant (ECCS-0708905), and by the DARPA QuEST program.

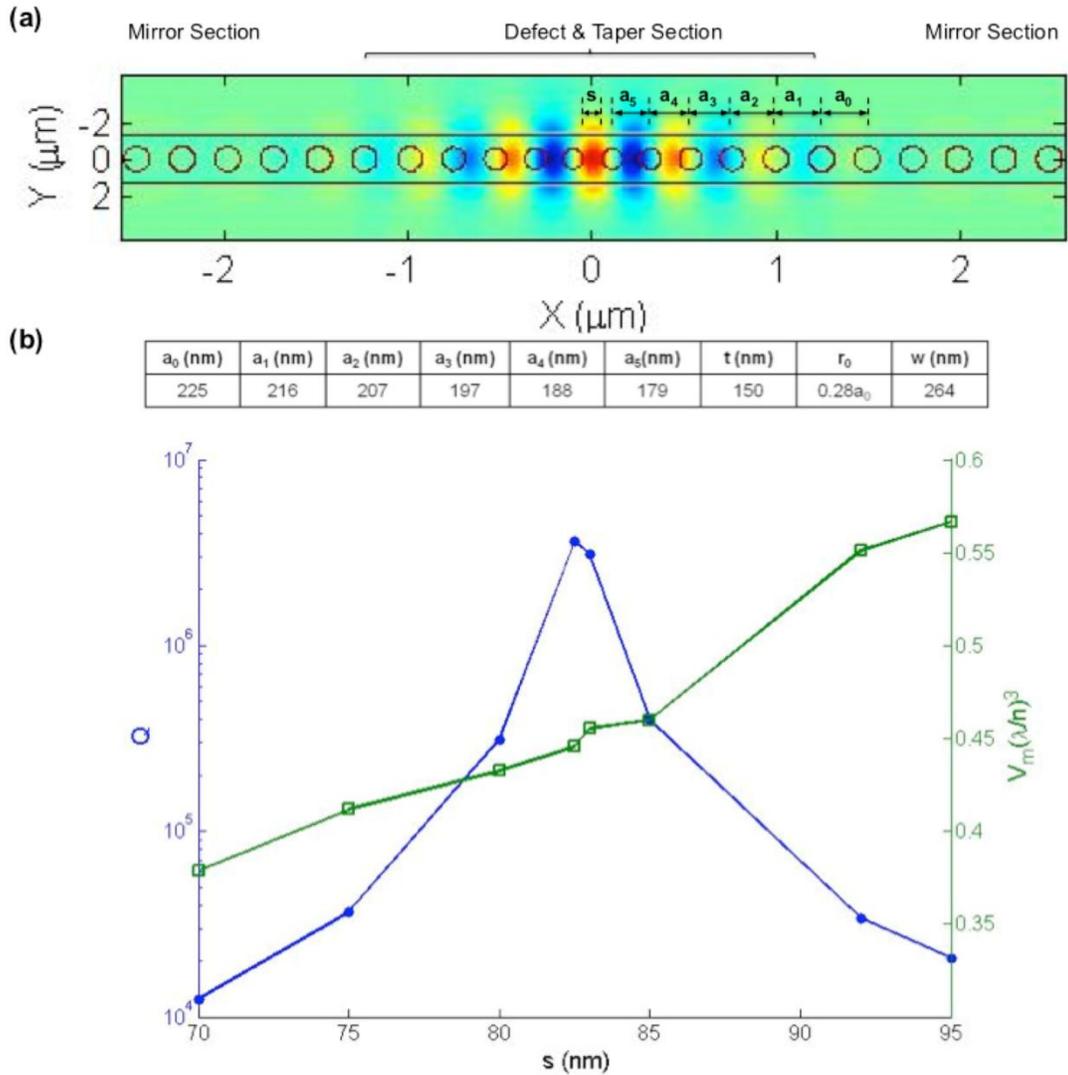

**Figure 1. (a)** Electric field ($E_y$) profile of a 5-taper photonic crystal nanobeam cavity with defect formed by a position shift *s* between the two center holes. The taper is characterized by a linear decrease in the periodicity of the air holes around the defect ($a_5 < a_4 < ... < a_0$). The taper and mirror regions are symmetric in the x direction about the cavity center. **(b)** Plot of Q and V as a function of cavity length *s* in a 5-taper cavity with 15 mirror pairs. Cavity quality factor Q can vary by several orders of magnitude with only slight variations in the mode volume V. Inset shows a table of the device parameters used in the simulation.

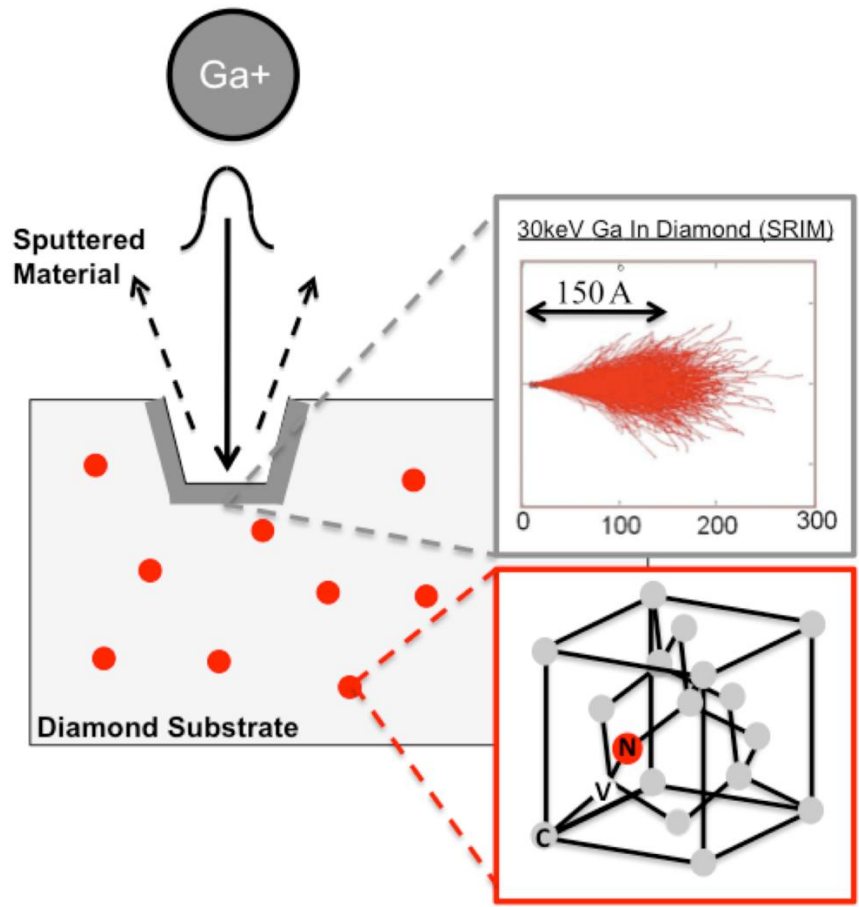

**Figure 2.** Focused ion beam (FIB) milling of structures in diamond containing Nitrogen-Vacancy (NV) centers. The FIB accelerates massive gallium ions towards the diamond substrate and sputters carbon material in a physical milling process. The width of the gallium ion beam – and hence resolution of the corresponding features – is strongly dependent upon the gallium beam current with optimal resolution ~5nm at 1pA/cm$^2$. NV centers distributed throughout the bulk are randomly embedded in fabricated structures. Grey insert shows the result of a stopping range of ions in matter (SRIM 2008[48]) calculation describing the penetration of gallium ions into diamond, which indicates that a thin layer of Gallium is implanted ~15nm beneath the surface. Red inset shows a cartoon of the NV color center in diamond, which is composed of an adjacent substitutional Nitrogen atom and vacancy.

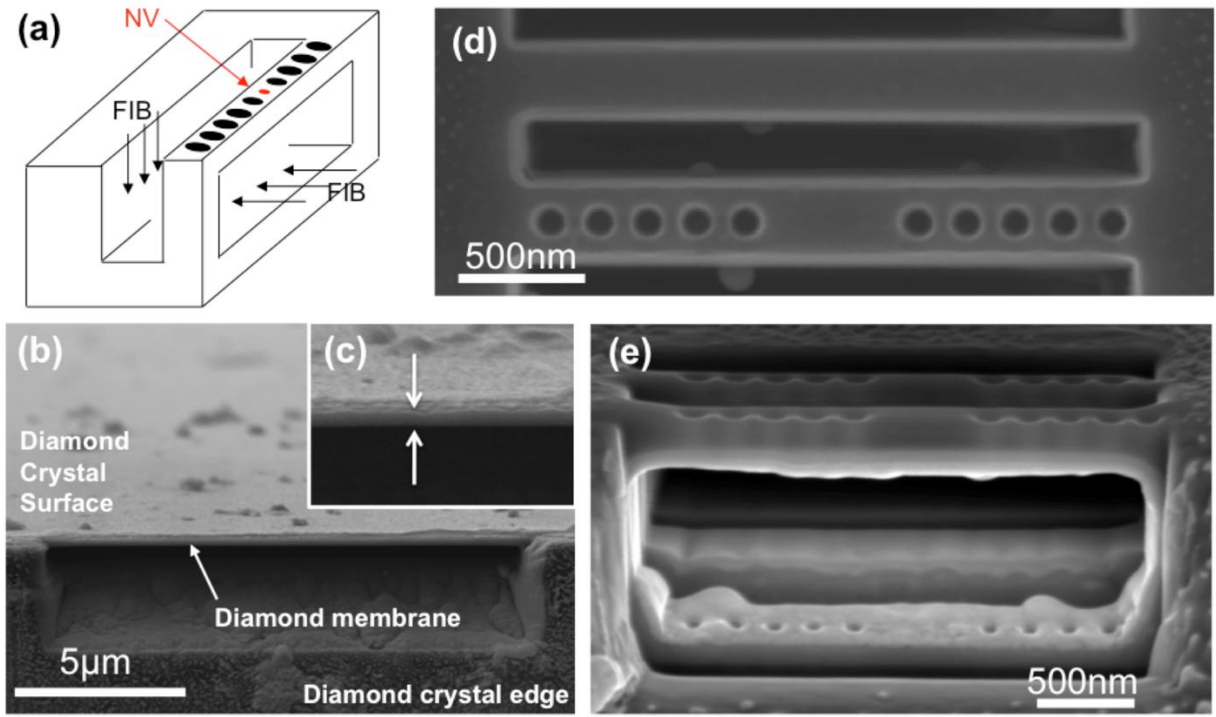

**Figure 3:** **(a)** Three-dimensional FIB sculpting procedure used to fabricate a nanobeam photonic crystal from a bulk diamond crystal based on crystal 'side-milling'. NV center would ideally be embedded in the defect region of the device. **(b)** SEM image of the side-view of a typical diamond membrane released from the bulk after milling a large pit in the side of the crystal. **(c)** High magnification SEM image of the membrane shown in part (b). The membrane is approximately 200nm thick (highlighted with white arrows). **(d)** SEM image shows a top-down view of a membrane after patterning with a waveguide and nanobeam photonic crystal cavity. Spots are residual carbon material re-sputtered on the device. **(e)** SEM image of the nanobeam resonators shows re-deposited amorphous carbon material. Holes observed in the bottom of the pit indicate that the FIB has milled through the diamond membrane completely.

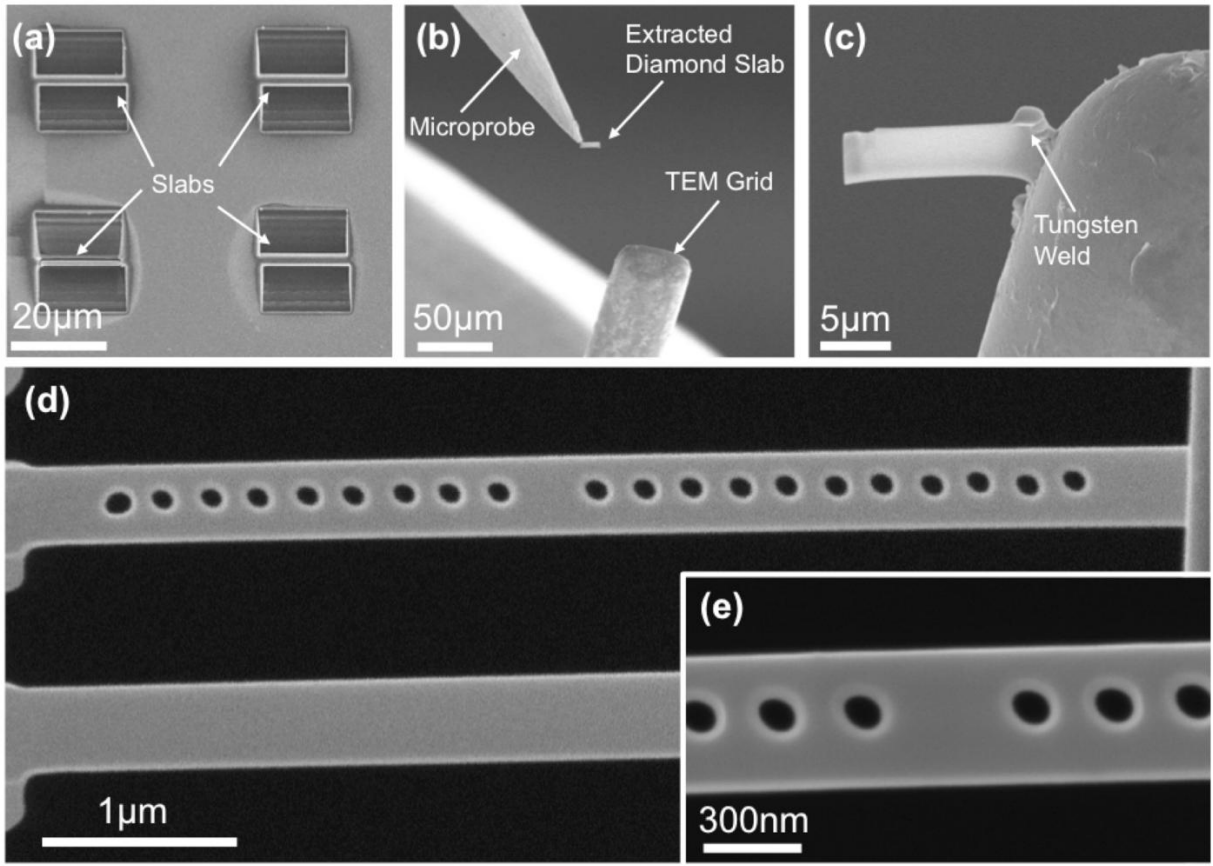

**Figure 4:** Three-dimensional FIB milling approach used to fabricate linear photonic crystals based on the TEM 'lift-off' technique. **(a)** SEM image of thick (~1-3μm) diamond slabs that were milled into the bulk crystal using ~300pA/cm$^2$ beam current. A microprobe is connected to a slab by ion beam assisted deposition of Tungsten, and the slab is released from the diamond crystal with milling steps on its left, right, and bottom sides (not shown). **(b)** SEM image of a released diamond membrane in the process of being transferred to a TEM grid. **(c)** SEM image of a diamond membrane that has been welded to a TEM grid. **(d)** SEM image of a thinned slab after patterning with waveguides and a photonic crystal nanobeam cavity. **(e)** High-magnification SEM image of the diamond nanobeam defect region.